\documentclass{jfm}
\usepackage{graphicx,colortbl,subcaption} 
\usepackage[normalem]{ulem} 
\usepackage{amsmath,amssymb}
\usepackage{pdfpages}
\usepackage{natbib}
\usepackage{booktabs}
\usepackage{gensymb}
\usepackage{comment}
\usepackage{tikz}

\usepackage[toc,page]{appendix}
\usepackage{newtxtext}
\usepackage{newtxmath}
\usepackage{hyperref}
\usepackage[colorinlistoftodos,textsize=tiny]{todonotes}

\hypersetup{
    colorlinks = true,
    urlcolor   = blue,
    citecolor  = black,
}

\newcommand{\RomanNumeralCaps}[1]
\linenumbers

\newcommand{\apremark}[1]{\textcolor{orange}{#1}}

\title{Geometry-controlled heat transport in differentially heated cavity through the lens of DNS and resolvent analysis}

\author{K. Chand \aff{1} 
\corresp{\email{krishan.iitghy@gmail.com},}, 
M. Quan\aff{1}, A. Padovan\aff{2},
 \and H. Luo\aff{1}
 }

\affiliation{
\aff{1} Department of Mechanical Engineering, Vanderbilt University, Nashville, TN (37209), USA

\aff{2}Department of Mechanical and Industrial Engineering, New
Jersey Institute of Technology, Newark, NJ 07102, USA
}

\begin{document}
\maketitle

\begin{abstract}
We perform direct numerical simulations (DNS) and resolvent analysis of natural convection in a differentially heated cavity (DHC) at Prandtl number $Pr = 0.7$, systematically varying the aspect ratio $\Gamma$ over $0.1 \leq \Gamma \leq 60$ and Rayleigh number $Ra = 10^5$--$10^8$. Across this nearly three-decade range of $\Gamma$, the Nusselt number $Nu$ exhibits four distinct regimes driven purely by geometric confinement via altering the structure of large-scale circulation (LSC): $Nu$ rises in regime I, peaks and plateaus in regime II, falls monotonically in regime III, and reaches a minimum and plateaus again in regime IV. 
Resolvent analysis shows a corresponding shift in the flow's response structure across these regimes: the maximum gain drops monotonically through regime I, bottoms out in regime II, and rises sharply through regime III and IV. 
The dominant response mode also changes character---from a diagonal mode peaking near the adiabatic walls, to a horizontally oriented structure peaking in the bulk flow (regimes I--II), to a vertical stack of such structures in regimes III--IV. The resolvent response tracks the sensitivity of the LSC to perturbations, which is minimized in regime II and maximized in regimes III and IV. 
Notably, the LSC configuration with the \textit{lowest} sensitivity consistently corresponds to the \textit{highest} heat flux, across all $Ra$ examined. We link this sensitivity to the LSC's anisotropy, quantified by the ratio of horizontal-to-vertical velocity Reynolds numbers ($Re_u/Re_v$). Remarkably, this ratio stays nearly constant ($\approx 0.45$) at the optimal (heat-transport-maximizing) aspect ratio ($\Gamma_{\mathrm{opt}}$) across all $Ra$, and the corresponding $\Gamma_{\mathrm{opt}}$ follows the power-law scaling
$\Gamma_{\mathrm{opt}} \sim Ra^{-0.17 \pm 0.01}$.
\end{abstract}

\begin{keywords}
Vertical convection, Large-scale circulation, and Resolvent analysis 
\end{keywords}
\section{Introduction}
\label{sec:intro}

Vertical convection (VC) in DHC is a canonical configuration in fluid mechanics with an extensive literature \citep{bejan2013convection}. It underpins a broad range of engineering and geophysical applications, including building insulation, electronic cooling, oceanic overturning circulation, and crystal-growth processes \citep{Ostrach1988,Ahlers_2009rev}. 
Unlike Rayleigh--B\'enard convection (RBC), vertical convection lacks exact global dissipation identities of the Grossmann--Lohse type~\citep{GL_2000,Ng_2015,shishkina_momentum_2016}, making the predictive description of heat transport substantially more challenging. In VC, while the dependence of the $Nu$ on $Ra$ and $Pr$ has been extensively studied, how geometric confinement reorganizes transport pathways remains poorly understood. 
For example, how the aspect ratio $\Gamma$ influences $Nu$ has not yet been mapped across the broad confinement range that is needed to reveal regime transitions and optimal transport states.

The first systematic theoretical treatment of convection in a rectangular cavity with differentially heated end walls was given by \citet{gill_boundary-layer_1966}, who derived boundary-layer solutions for tall cavities ($\Gamma \gg 1$). \citet{Bejan1979} refined Gill’s analysis by enforcing zero net upward energy flux at the horizontal boundaries, yielding improved $Nu$ correlations that depend on both $Ra$ and $\Gamma$. Subsequent asymptotic and numerical studies established the limiting small- and large-$\Gamma$ behavior \citep{Cormack1974a,Cormack1974b,macgregor_free_1969}, including the widely used correlation, $Nu = 0.42 \Gamma^{-0.30} Pr^{0.012} Ra^{0.25}$, for air-filled cavities \citep{macgregor_free_1969}. Despite these important advances, most modern studies of vertical convection focus on $Nu(Ra,Pr)$ and $Re(Ra,Pr)$ at fixed geometries, leaving the transport consequences of systematic variation in $\Gamma$ still unresolved.

Beyond heat-transfer correlations, the aspect ratio strongly influences underlying flow organizations and the route to turbulence. Previous studies linked corner turning and transition dynamics to hydraulic-jump and thermal mechanisms \citep{Paolucci_Chenoweth_1989,Ravi_1994}, while the broader literature on vertical convection has focused on laminar and turbulent scaling with $Ra$ and $Pr$~\citep{Ng_2015,shishkina_momentum_2016,Howland2022}. However, a mechanistic framework connecting the aspect-ratio variation to large-scale circulation (LSC), instability pathways, and heat transport optimization is still lacking.


In the present work, we show that varying $\Gamma$ over nearly three orders of magnitude reorganizes the LSC into four distinct transport regimes, each characterized by a qualitatively different LSC. 
To understand the underlying mechanism, we combine direct numerical simulations with resolvent analysis and show that the regime transitions coincide with maximum resolvent gain-aspect ratio curve and the corresponding qualitative changes in the dominant response structure.  The present study focuses on air ($Pr=0.7$) and employs two-dimensional simulations over a wide $(Ra,\, \Gamma)$ parameter space. This choice is physically motivated by the quasi-two-dimensional character of vertical convection over a substantially broader range of $Ra$ than RBC, owing to mean shear that suppresses spanwise instabilities~\citep{Ng_2015,Wang_2021}. Prior comparative studies at comparable $Ra$ and $Pr$ further show close agreement between two- and three-dimensional results in both $Nu$ and LSC structures~\citep{LeQuere1998,chand_effect_2022}. Moreover, a systematic sweep over a large range of $\Gamma$ across multiple $Ra$ would be computationally prohibitive in three dimensions for both DNS and resolvent analysis. Therefore, the present two-dimensional framework enables the identification of the geometry-controlled heat transport mechanism that govern $Nu(\Gamma)$ regimes.

\begin{figure}
            \centering
            \includegraphics[width=\linewidth]{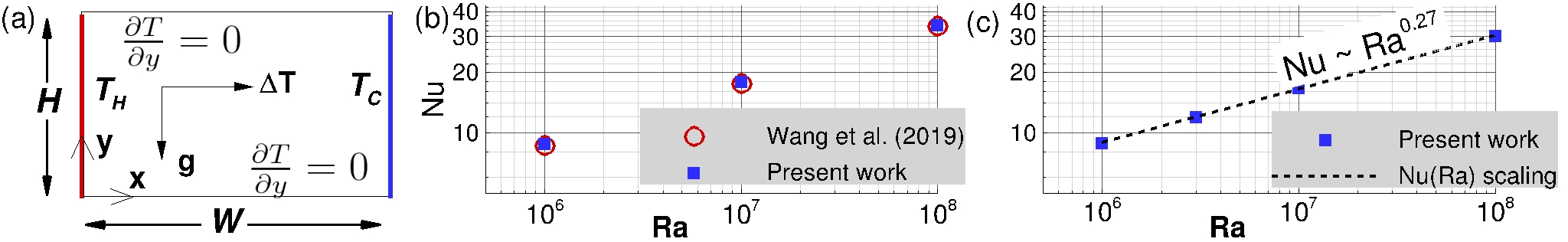}
            
     \caption{(a) Schematic showing the flow domain and boundary conditions. (b) Validation of the flow solver with \citet{Wang_2019} for $\Gamma=0.5$ and $Pr=0.7$. (c) Scaling of $Nu$ with $Ra$ for $\Gamma=1$ and $Pr=0.7$.}
    \label{fig:schematic}
\end{figure}

\section{Mathematical model}

\subsection{Direct numerical simulation}
The model setup is shown in Fig.~\ref{fig:schematic}(a), where all sides of the domain are no-slip and no-penetration walls. The left and right sides are at specified hot and cold temperatures, 
$T_H$ and $T_C$, respectively. The top and bottom walls are insulated.
The governing equations are the viscous incompressible Navier-Stokes equation  coupled with the energy equation under the Boussinesq approximation. Their non-dimensionalized form is given by: 
\begin{eqnarray}\label{eq:1}
\partial_j u_j &=& 0\\
\partial_t u_i + u_j \partial_j u_i &=& -\partial_i p + \sqrt{Pr/Ra}\, \partial^2_j u_i + \theta \delta_{i2}\\ \label{eq:3}
\partial_t \theta + u_j \partial_j \theta &=& 1/\sqrt{PrRa} \, \partial^2_{jj} \theta
\end{eqnarray}
where $i=1$ and 2, representing the Cartesian coordinates $x$ and $y$, respectively, $t$ is time, $u_i = (u, v)$ is the velocity, $p$ is the pressure, and $\theta = (T - T_C)/(T_H - T_C)$ is the non-dimensional temperature and $\delta_{i2}$ is Kroneckar delta. Here, $Ra = g \beta \Delta T W^3 / (\nu \alpha)$ and $\Pran = \nu / \alpha$, where $g$ is the gravitational acceleration, $W$ is width of the domain, $\beta$, $\nu$, and $\alpha$ denote the thermal expansion coefficient, kinematic viscosity, and thermal diffusivity, respectively. For non-dimensionalization, we use $W$, the free-fall velocity $U = \sqrt{g \beta \Delta T W}$, and temperature difference $\Delta T = T_H - T_C$ as length, velocity, and temperature scales, respectively. 

In Fig.~\ref{fig:schematic}(b), we show validation results against \citet{Wang_2019} for $\Gamma=0.5$. 
We also show $Nu(Ra)$ scaling in Fig.~\ref{fig:schematic}(c) for $\Gamma = 1$ and $10^6 \leq Ra \leq 10^8$, where $Nu \sim Ra^{0.27}$ is consistent with \citep{Ng_2015,Wang_2021}. Following the work of \citet{Wang_2021} for $\Gamma=1$, we scale the grid points accordingly to ensure that both the Kolmogorov length scale and boundary layers are fully resolved. Numerical details of all cases are tabulated in Tables~\ref{tab:1} and \ref{tab:2}, see appendix \ref{app:1}. This solver is extensively used in our previous works \citep{chand_effect_2022,sharma_dominant_2024}. 
In the analysis, we use the following definitions of Nusselt and Reynolds numbers $(Re)$, 
\begin{equation}
Nu = \langle |\nabla \theta|^2 \rangle, 
\hspace{5pt}
Re = \sqrt{Ra\langle u_i u_i \rangle/Pr},
\hspace{5pt}
Re_u = \sqrt{Ra/Pr\,\langle u^2\rangle},
\hspace{5pt}
Re_v = \sqrt{Ra/Pr\,\langle v^2\rangle},
\label{eq:Nu}
\end{equation}
where $\langle \cdot \rangle$ denotes volume- and time-averaging (for unsteady cases). 
The ratio $Re_u/Re_v$ quantifies the anisotropy in the characteristic of LSC. 
Note that $\langle \cdot \rangle_t$ is also used later to represent a time-averaged quantity.

\subsection{Resolvent analysis}
Resolvent analysis provides the input-output framework to study amplification of disturbances in  flows, where the nonlinear terms of the Navier--Stokes equations are treated as an endogenous forcing acting on the linearized dynamics about a base state. \citet{jovanovic_componentwise_2005} laid its systems-theoretic foundations and showed that the frequency response selectively amplifies stochastically forced disturbances in a channel flow, which resembles with the near-wall streaks and streamwise vortices. Later, \citet{mckeon_critical-layer_2010} extended this analysis to turbulent pipe flow to illuminate the scaling behavior of very large-scale motions. For further details on resolvent analysis, see \citet{rolandi_invitation_2024}. 
Here, we extend this resolvent-based perspective to the present buoyancy-driven flows, where the base flow is a steady large-scale circulation in majority of the cases except for $\Gamma \geq 5$, where a time-averaged flow field is considered as the base flow. 
 
By subtracting the base flow  equation from equations \eqref{eq:1}--\eqref{eq:3}, the dynamics of perturbation is governed by linearized Boussinesq equations:
\begin{eqnarray}\label{eq:res1}
\partial_j u'_j &=& 0\\
\quad \partial_t u'_i + u'_j \partial_j u_{b_i} + u_{b_j} \partial_j u'_i &=& -\partial_i p' + \sqrt{Pr/Ra}\, \partial^2_j u'_i + \theta' \delta_{i2} +f'_i\\ \label{eq:res3}
\partial_t \theta' + u'_j \partial_j \theta_b + u_{b_j} \partial_j \theta' &=& 1/\sqrt{PrRa}\, \partial^2_{jj} \theta' + f'_\theta
\end{eqnarray}
The nonlinear forcing terms can be related to the Reynolds-stresses: $f'_i = -\partial_j (u_i' u_j' -
\langle u'_i u'_j \rangle_t)$ and $f'_\theta= -\partial_j (u'_j \theta' -
\langle u'_j \theta' \rangle_t)$. Both the fluctuating velocity and temperature fields satisfy homogeneous Dirichlet conditions at the walls. For base state $q_b = [u_b, v_b, \theta_b, p_b]$ and perturbation $q' = [u', v', \theta', p']$, the linearized form of equations \ref{eq:res1}--\ref{eq:res3} along with control output $y$ can be written as 
\begin{equation}\label{eq:descriptor_system1} 
    M\frac{\partial q'}{\partial t} = Lq' + Mf', \qquad \text{and} \qquad
    y = C M q'
\end{equation}
where $C$ and $M$ are transfer functions, defined soon, $L$ is the linearized operator, 
\begin{equation} \label{eq:app_linearoperator}
\small
L =
\begin{bmatrix}
-  \mathbf{u}_b \cdot \nabla - \partial_x  u_b + \sqrt{\frac{Pr}{Ra}}\,\nabla^2 &
-\partial_y  u_b & 0 & -\partial_x \\
-\partial_x  v_b & -  \mathbf{u}_b \cdot \nabla - \partial_y  v_b + \sqrt{\frac{Pr}{Ra}}\,\nabla^2 & 1 & -\partial_y \\
-\partial_x  \theta_b & -\partial_y  \theta_b & -  \mathbf{u}_b \cdot \nabla
+ \frac{1}{\sqrt{Pr\,Ra}}\,\nabla^2 & 0 \\
\partial_x & \partial_y & 0 & 0
\end{bmatrix}.
\end{equation}
and $y = Nu' (=  \nabla \theta' \cdot \hat{n}_x)$, is the wall-normal $(\hat{n}_x)$ temperature-gradient fluctuation evaluated at the isothermal walls. 

Considering $q' = \hat{q}(x,y) e^{-i \omega t}$ and $f' = \hat{f}(x,y) e^{-i \omega t}$, where $\omega$ is frequency, equations \ref{eq:descriptor_system1} can be written as
\begin{align}
\small
    \hat{q} &= \left(-i\omega M - L\right)^{-1}M\hat{f}, \\
    \hat{y} = C M \left(-i\omega M - L\right)^{-1}M\hat{f}\quad \text{or} \quad \hat{y}&= CM\mathcal{H}(\omega)M\hat{f},
\end{align}
where $\mathcal{H}$ is the resolvent operator. All the operators ($C$, $M$, $L$ and $\mathcal{H}(\omega)$) are discretized spatially on a staggered grid with $N_u=(N_x+1) N_y$, $N_v=(N_y+1) N_x$, and $N_\theta=N_p=N_x N_y$, where $N_x$ and $N_y$ are the collocated grid points. The total grid points are: $N=N_u+N_v+2N_p$. Here, $M = \mathrm{diag}(I_{N_u}, I_{N_v}, I_{N_\theta}, 0)$ is a block-diagonal mass matrix, $C = [0_{N_u},0_{N_v},G_{\theta_{N_\theta}},0_{N_p}]$ is a block row operator matrix in which $G_\theta$ evaluates the wall-normal gradient at the isothermal walls. Note that $M$ eliminates pressure from the evolution equations to enforce discrete incompressibility. 

Singular value decomposition (SVD) of $\mathcal{H}(\omega)
    =
    \sum_{j=1}^{N}
    \hat{\psi}_{j}\sigma_j \hat{f}_{j}^{*}$,    
gives the gain, forcing, and response modes, where $\sigma_j$ $(\sigma_1 \geq \sigma_2 \geq \dots \geq \sigma_N)$ is the resolvent gains, $\hat{f}_{j}$ and $\hat{\psi}_{j}$ are the right and left singular vectors corresponding to forcing and response modes, respectively. 
The SVD of $\mathcal{H}(\omega)$ is evaluated in a norm of $L^2$. 
We restrict the analysis to $\omega=0$, which isolates the response of $Nu'$ to steady forcing. We use \texttt{Resolvent4py}, an open-source Python package \citep{padovan_resolvent4py_2025}, to conduct resolvent analysis.
\section{Results and discussion}\label{sec:results}
\begin{figure}
\centering
\includegraphics[width=0.9\textwidth]{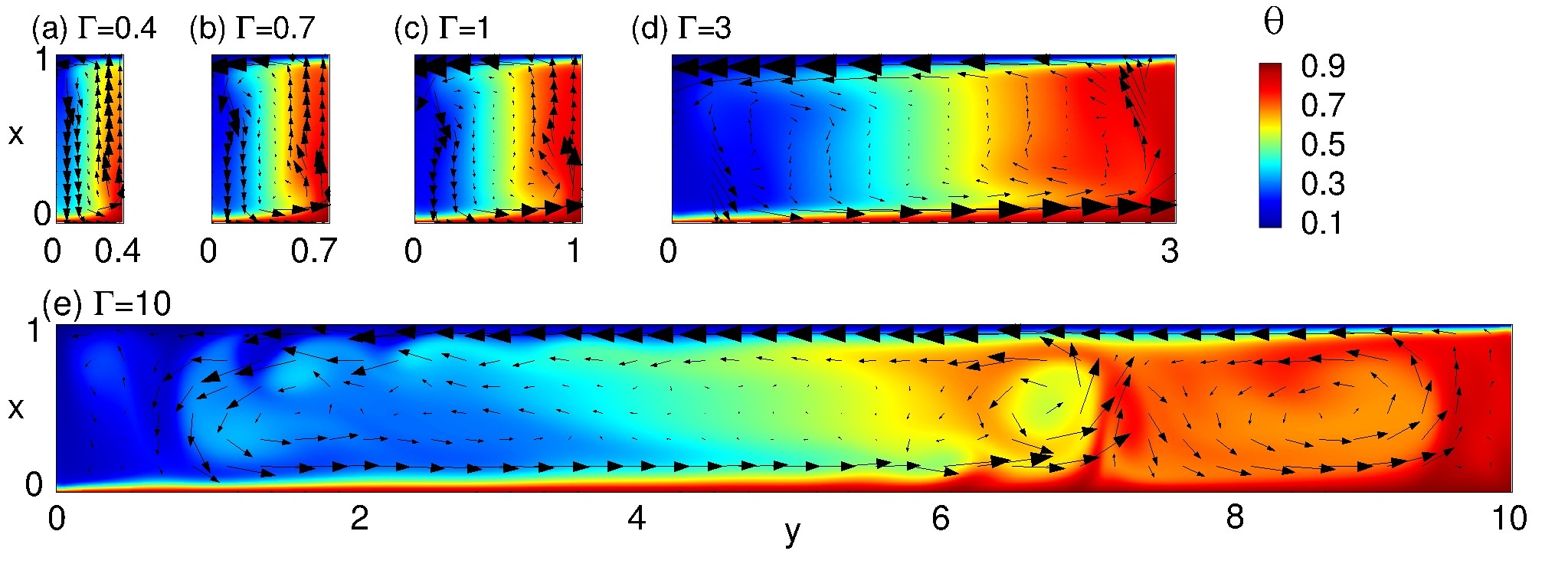}
\caption{For $Ra=10^6$, instantaneous temperature contours overlaid with velocity vectors, illustrating the geometry-induced reorganization of the large-scale circulation. For a better representation, we show the visualizations horizontally ($y$ is the vertical direction). }
\label{fig:instantfield}
\end{figure}

We begin by examining how the organization of the LSC changes with aspect ratio, as shown in Fig.~\ref{fig:instantfield}, at $Ra=10^6$. At $\Gamma=0.4$, the LSC is predominantly horizontal (aligned in $x$), whereas at $\Gamma=0.7$, it remains primarily horizontal but with noticeable vertical movement (in $y$-axis). At $\Gamma=1$, the vertical movement of the flow slightly increases and leads to a squarish shape, rather than remaining purely horizontally dominated. At $\Gamma=3$, the flow becomes predominantly vertical, with only weak horizontal movement. Finally, at $\Gamma=10$, we observe oscillatory flow behavior, and a single LSC breaks down into multiple rolls but remains predominantly vertical. This shows that $\Gamma$ alters the the LSC (from steady laminar to unsteady flow), which in turn impacts heat transfer, as LSC plays a central role in this process. In what follows, we examine how $Nu$ depends on the LSC and $\Gamma$ across different $Ra$, and we characterize this dependence in terms of distinct regimes associated with LSC.

\subsection{Regime identifications and resolvent analysis
}\label{sec:results1}
\begin{figure}\centering
\begin{subfigure}{0.495\textwidth}
\includegraphics[width=\textwidth]{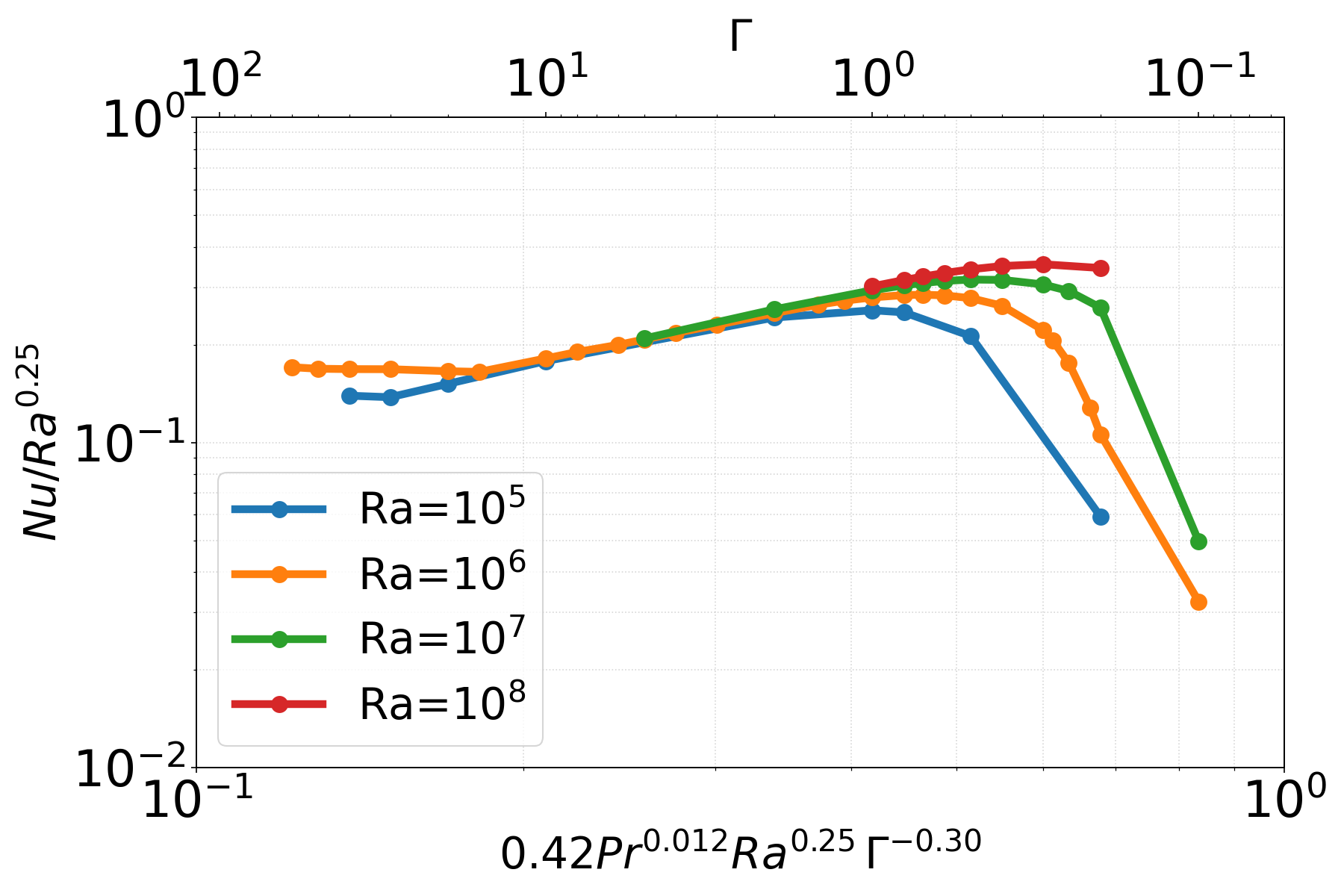}
\caption{}\label{fig:1a}
\end{subfigure}
\begin{subfigure}{0.495\textwidth}
\includegraphics[width=\textwidth]{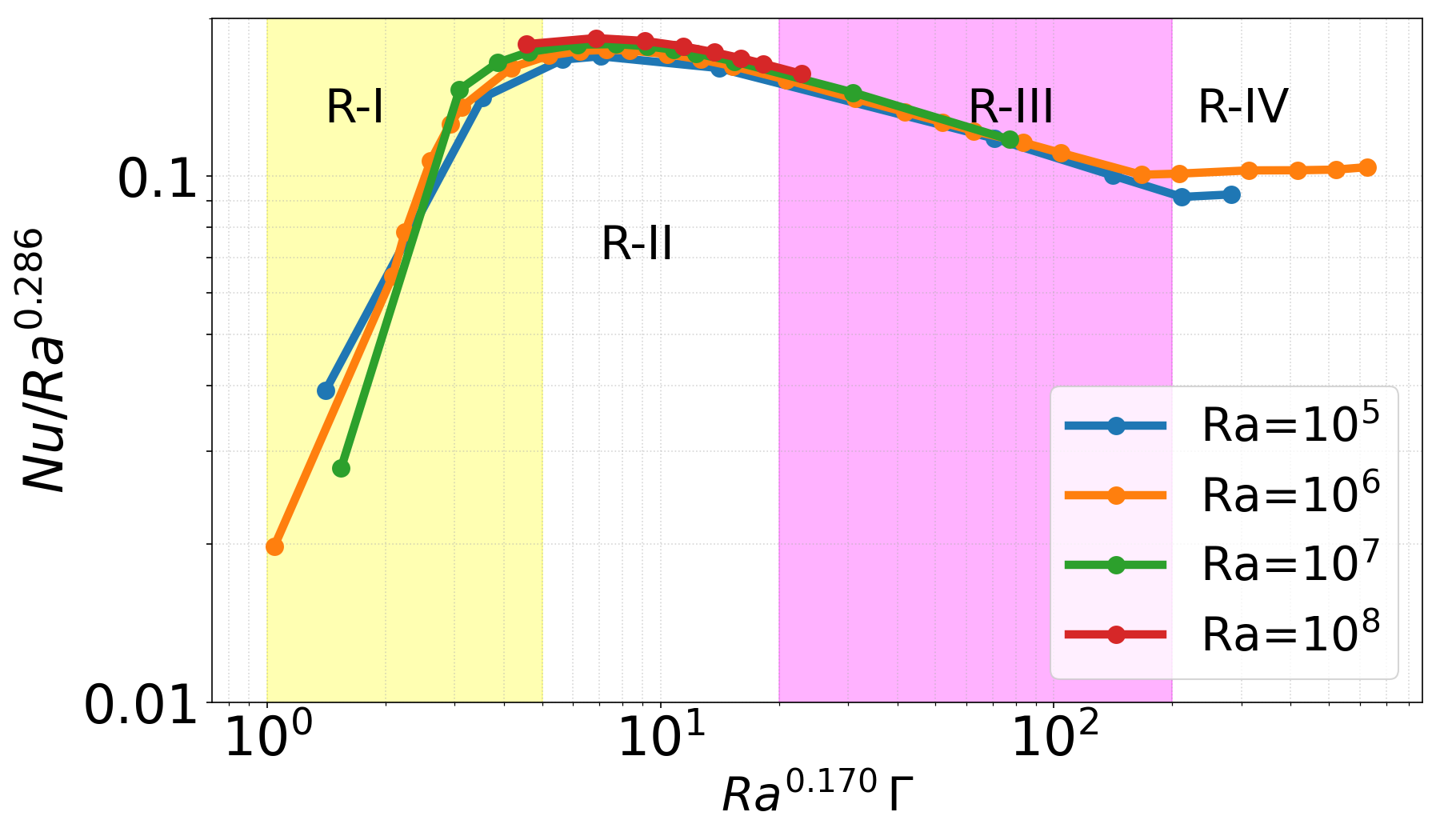}
\caption{}\label{fig:1b}
\end{subfigure}
\begin{subfigure}{0.495\textwidth}
\includegraphics[width=\textwidth]{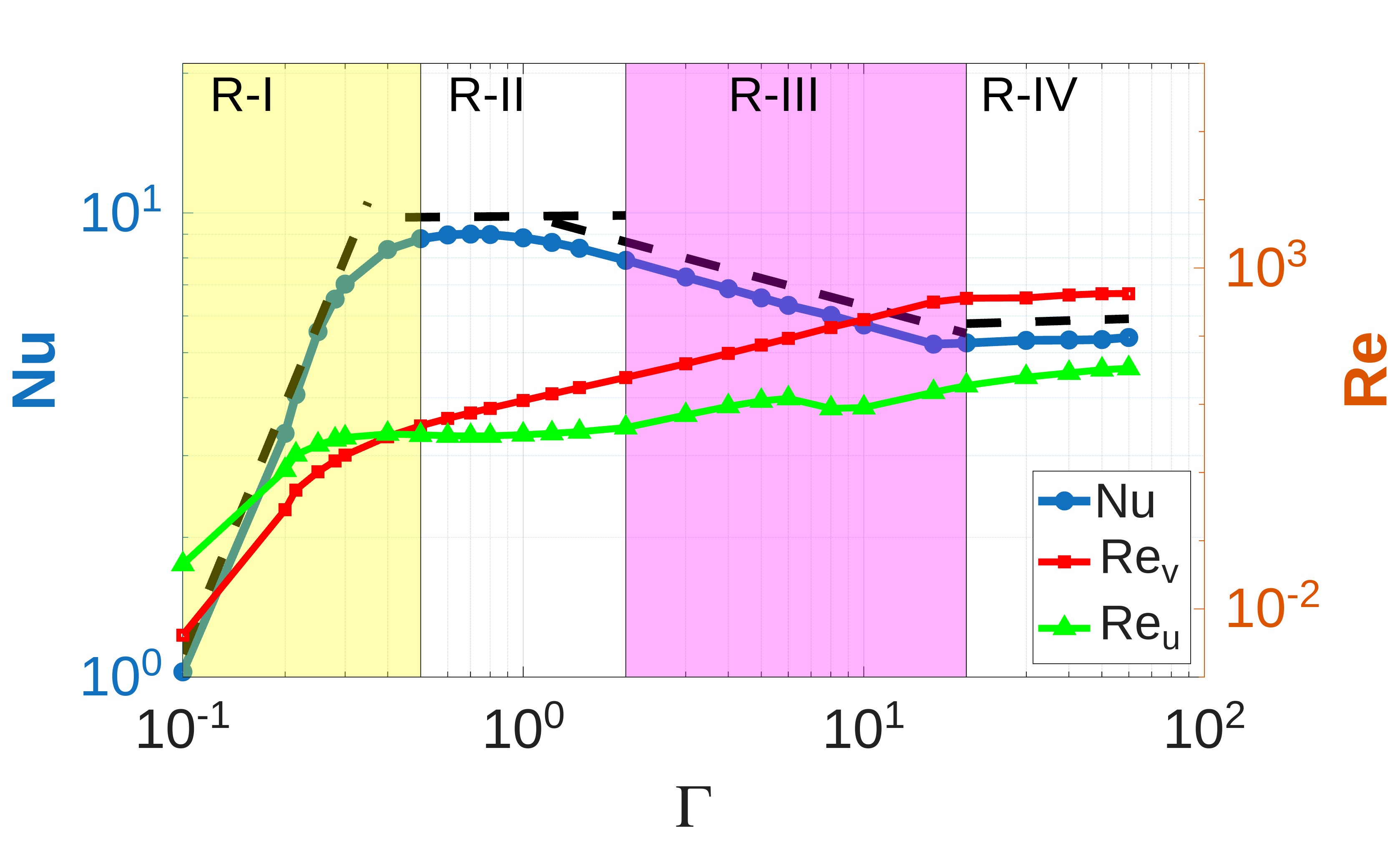}
\caption{}\label{fig:1c}
\end{subfigure}
\begin{subfigure}{0.495\textwidth}
\includegraphics[width=\textwidth]{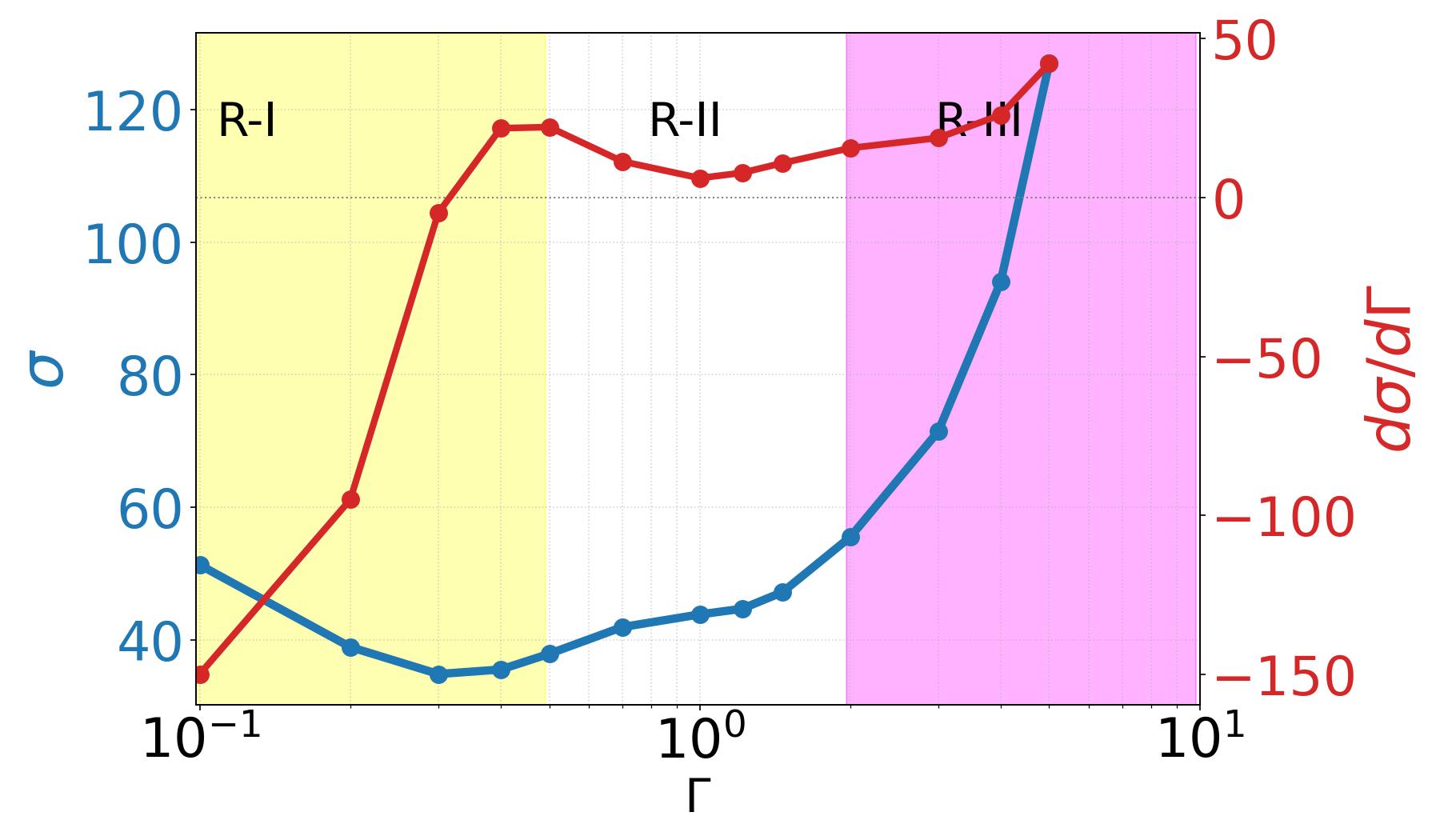}
\caption{}\label{fig:1d}
\end{subfigure}


\caption{(a) $Nu(Pr Ra \Gamma)$ comparison with \citet{macgregor_free_1969},  with an additional top x-axis showing $\Gamma$. (b) $Nu(\Gamma Ra)$ showing the collapse across the $Ra$ and the four regimes; (c) $Nu$, $Re_u$, and $Re_v$ plotted against $\Gamma$ at $Ra=10^6$, where dashed lines are scaling references given by $\Gamma^b$, $b=1.81\pm 0.05$, $0.01\pm 0.02$, $0.20\pm0.0$, and $0.02\pm0.01$, for the four regimes, respectively. (d) Maximum resolvent gain ($\sigma)$ and its slope ($d\sigma/d\Gamma)$ vs $\Gamma$ for the cases in (c).  
}
\label{fig:Nuregime}
\end{figure}



Figure~\ref{fig:1a} shows the dependence of scaled $Nu$ on $\Gamma$ for different $Ra$ cases. All $Ra$ cases collapse onto each other in the $\Gamma$ range $1$ to $20$ (see the top x-axis), which is consistent with the correlation given by \citet{macgregor_free_1969}. 
However, the trend clearly shows that the correlation deviates at both extremes of $\Gamma$, specifically at $\Gamma > 20$ and $\Gamma < 1$, for a fixed $Pr=0.7$. 
In Fig.~\ref{fig:1b}, we show another rescaled plot of $Nu$ versus $\Gamma$ to establish that the four regimes are observed across all $Ra$ cases. From regime R-I to R-IV, $Nu$ first increases, then remains nearly constant, then monotonically decreases, and finally saturates at a lower $Nu$ in R-IV. Since these four regimes are consistent across all $Ra$ cases, we fix $Ra=10^6$ to isolate the sole effect of $\Gamma$ (see Fig.~\ref{fig:1c}). In these regimes, we also quantify the characteristics of the LSC through the Reynolds-number components defined in Eq.~\ref{eq:Nu}. The ratio of $Re_u$ and $Re_v$ quantifies the LSC's dominant direction and hence measures its geometric anisotropy. 

\textit{Regime~I: Low-ceiling regime ($\Gamma \lesssim 0.5,\; Nu \sim \Gamma^{1.81\pm0.05}$)}. In this regime, 
$Nu$ increases sharply with $\Gamma$ from the stagnant state at $\Gamma = 0.1$, where strong vertical confinement inhibits fluid motion, owing to the absence of convective transport. The resulting LSC structure is horizontally dominated with $Re_u > Re_v$.

\textit{Regime~II: Optimal transport regime} ($0.5 \lesssim \Gamma \lesssim 2,\; Nu \sim \Gamma^{0.01\pm0.02}$).  
In this regime, $Nu$ plateaus and attains its maximum value. Interestingly, $Re_v$ crosses over $Re_u$ because $Re_v$ continues to grow, while $Re_u$ remains nearly constant throughout the regime. Here, the circulation is reshaped into a dynamically balanced and stable LSC. At $\Gamma=0.7$, optimum heat transfer occurs under a certain geometrical anisotropy of the LSC, i.e.  $Re_u/Re_v \approx 0.45$. Although $Re_v$ increases in this regime, the ratio $Re_u/Re_v$ remains close to $0.45$. For varying $Ra$, we find that maximum $Nu$ always occurs when $Re_u/Re_v \approx 0.45$, as discussed later in the next section.

\textit{Regime~III: Weak-transport regime} ($2 \leq \Gamma \leq 20,\; Nu \sim \Gamma^{-0.20\pm0.01}$).  
A further increase in $\Gamma$ leads to a well-studied regime, where the correlation of \citet{macgregor_free_1969} works well for the $Nu(\Gamma Pr Ra)$ dependence. In this regime, the LSC elongates in the vertical direction with increasing $\Gamma$, resulting in a monotonically increasing $Re_v$ and a slight increase in $Re_u$, which leads to $Re_u/Re_v \lesssim 0.1$. This vertically biased characteristic of the LSC weakens horizontal heat transport and promotes oscillations. Within this regime, the LSC changes from a stable, vertically dominated structure at $\Gamma \leq 4$ to a state of multiple rolls at larger $\Gamma \geq 10$, as shown in Fig.~\ref{fig:instantfield}e. The number of such rolls increases with $\Gamma$.

\textit{Regime~IV: Asymptotic channel regime} ($\Gamma \gtrsim 20,\; Nu \sim \Gamma^{0.02\pm0.01}$). At 
a sufficiently large aspect ratio, the influence of vertical confinement vanishes, so that $Nu$, $Re_u$, and $Re_v$ become independent of $\Gamma$. The anisotropy of the LSC also saturates  at $Re_u/Re_v \approx 0.07$. As in R-III, a vertical stack of rolls is present throughout the regime, and the number of these rolls increases with $\Gamma$.

The properties of $Nu$ and $Re$ across the four regimes show that neither $Re_u$ nor $Re_v$ alone controls the heat transfer, but their combined effect does, which reflects the structure of the LSC. To understand the sensitivity of the flow to perturbations for optimum heat transfer, we perform resolvent analysis under steady forcing ($\omega=0$). Figure \ref{fig:1d} shows the maximum resolvent gain versus $\Gamma$, together with the slope of the $\sigma$--$\Gamma$ curve. The gain curve is consistent with the $Nu$--$\Gamma$ regimes identified above (only three regimes are shown, owing to the significantly larger $\sigma$ in R-IV). The curve indicates that the peak gain decreases with $\Gamma$ to reach a trough in R-II, before increasing sharply in R-III and R-IV. Note that the slope of the $\sigma$--$\Gamma$ curve approaches and stays close to zero in R-II and departs from zero elsewhere. The minimum gain corresponds to the flow (LSC) that yields the maximum $Nu$. This implies that the LSC with $Re_u/Re_v \approx 0.45$ exhibits the least amplification of fluctuations. In R-III, however, the sudden rise in gain indicates that the flow is more sensitive to perturbations. These results are consistent with the oscillatory flow structures observed in Fig.~\ref{fig:instantfield}e and prior observations that the flow becomes unsteady and exhibit oscillatory behavior at larger $\Gamma$ \citep{Gill1969}.
\begin{figure}
\centering
\includegraphics[width=0.9\textwidth]{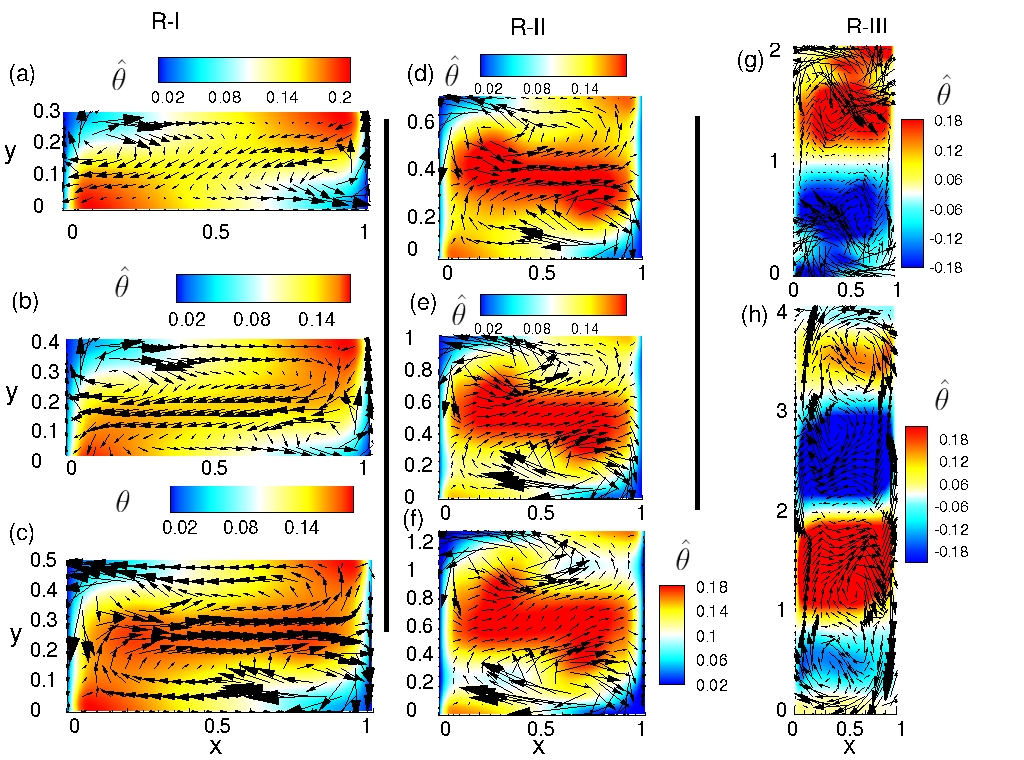}
\caption{Response mode $\hat{q}$ at $\omega=0$, reconstructed from the optimal forcing mode $\hat{f}_1$, for (a) $\Gamma=0.3$, (b) $0.4$, (c) $0.5$, (d) $0.7$, (e) $1$, (f) $1.215$, (g) $2$ and (h) $4$. Panels (a)--(c) lie in R-I, (d)--(f) in R-II and (g)--(h) in R-III. Contours show $\hat{\theta}$, overlaid with velocity vectors. 
}
\label{fig:resolventmodes}
\end{figure}


Across the three regimes in Fig.~\ref{fig:1d}, we further show the corresponding most amplified response modes in Fig.~\ref{fig:resolventmodes}. In R-I $(\Gamma < 0.5)$, the peak response of temperature fluctuation resides at the adiabatic walls and leads to a diagonally dominant structure  and transitioning at $\Gamma=0.5$ (panels a-c). With increasing $\Gamma$, the peak response shifts from the adiabatic walls to the bulk (panels d-f). In R-III, the peak response indicates vertically stacked rolls. Similar vertically stacked rolls are also observed in R-IV, wherein the number of rolls increases with $\Gamma$. The analysis suggests that a stable LSC with peak response in the bulk is less responsive to perturbations and leads to optimum heat transport. 
\subsection{Optimum heat transfer}
\begin{figure}\centering
\begin{subfigure}{0.42\textwidth}
\includegraphics[width=\textwidth]{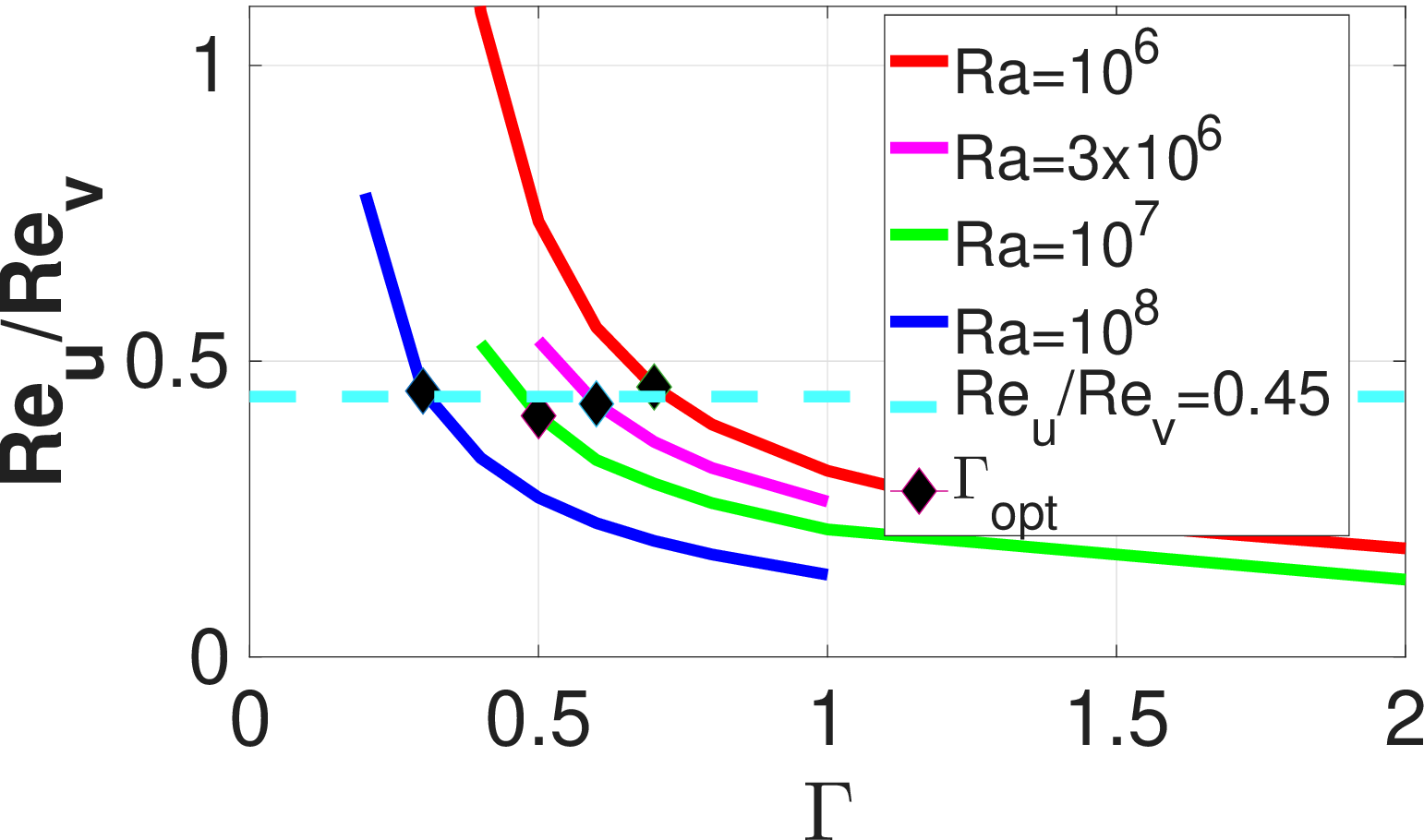}
\caption{}
\end{subfigure}
\begin{subfigure}{0.42\textwidth}
\includegraphics[width=\textwidth]{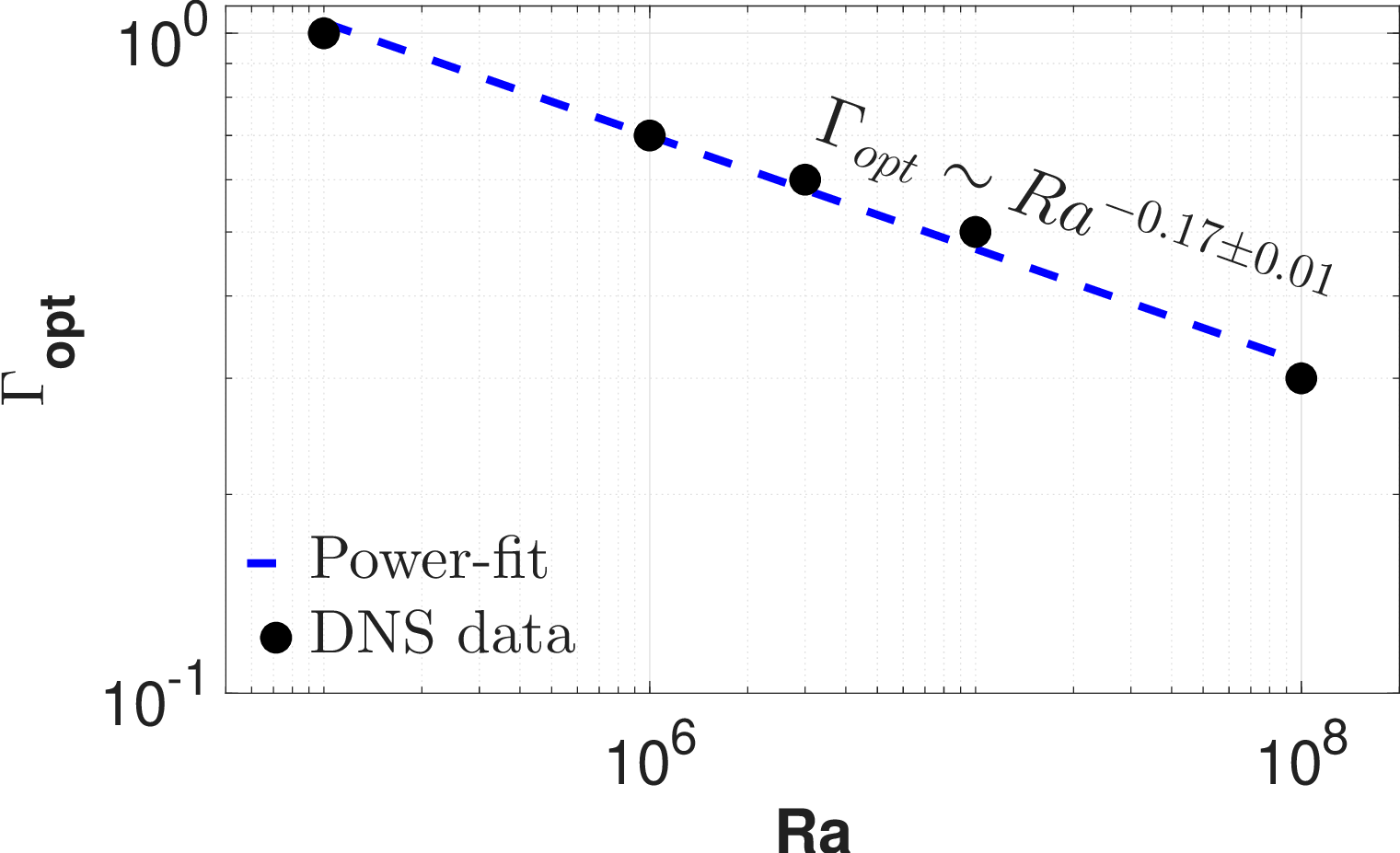}
\caption{}
\end{subfigure}
\caption{(a) Variation of the anisotropy ratio $Re_u/Re_v$ with $\Gamma$ across different $Ra$, 
and (b) scaling of the optimal aspect ratio $\Gamma_{\mathrm{opt}}$ with $Ra$, demonstrating the predictive geometry law for the optimal $Nu$.}
\label{fig:2}
\end{figure}

A key outcome of this geometry sweep is the identification of an optimal transport pattern. Here, $\Gamma_{\mathrm{opt}}$ is defined as the $\Gamma$ that gives maximum heat transfer for each $Ra$. Figure~\ref{fig:2}a shows that the peak heat transfer occurs when $Re_u/Re_v \approx 0.45$ (see cyan dashed line) across all $Ra$ considered, indicating the $Ra$-independent condition. 
The ratio signifies a balance between horizontal heat-carrying motion, which contributes to the heat transfer, and vertical buoyancy flow that triggers instability.  
For $Re_u/Re_v \gg 0.45$, the cavity is over-confined and the circulation remains excessively horizontal, limiting the vertical flow. On the other hand, for $Re_u/Re_v \ll 0.45$, the flow becomes overly elongated and susceptible to oscillations, which leads to the formation of multiple rolls and degrades the horizontal transport to diminish the overall heat flux. 

In Fig.~\ref{fig:2}b, we further show that the corresponding optimal geometry follows a clean power-law: $\Gamma_{\mathrm{opt}} \sim Ra^{-0.17\pm0.01}$. The power-law signifies the limit of efficient heat transfer through steady flow conditions at small $Ra$ for optimum heat transfer. At larger $Ra$, flow tends to become turbulent, and attaining a steady state would require to diminish $\Gamma$ to impractical situations, making it of no practical use. 
Hence, the scaling law provides a predictive law for geometry-driven heat-transfer optimization.

\section{Conclusions}\label{sec:conclusions}
Combining DNS at $Ra=10^5$--$10^8$ ($Pr = 0.7$) with an aspect ratio ($\Gamma$) sweep spanning nearly three orders of magnitude, we have shown that geometric confinement alone produces flow states ranging from stagnant to steady laminar to unsteady. We have found that $\Gamma$ reorganizes the large-scale circulation (LSC), and hence the dominant heat-transfer pathways, into a non-monotonic four-regime $Nu(\Gamma)$ dependence. 

The central physical mechanism is the aspect-ratio-induced restructuring of the LSC. In R-I ($\Gamma \lesssim 0.5$), strong confinement forces a horizontally dominant LSC that aligns with the imposed thermal gradient, resulting in the steep aspect ratio dependence, $Nu \sim \Gamma^{1.81}$. This horizontal dominance is quantified by the Reynolds numbers based on the horizontal and vertical velocities ($Re_u \gg Re_v$). In R-II ($0.5 \lesssim \Gamma \lesssim 2$), the circulation reorganizes into a dynamically balanced and transport-efficient structure ($Re_u/Re_v \approx 0.45$), producing a near-plateau and the maximum heat transport. In R-III ($2 \lesssim \Gamma \lesssim 20$), the LSC progressively elongates in the vertical direction ($Re_u/Re_v < 0.1$) and breaks into multiple rolls at large $\Gamma$, yielding the gradual decay with $Nu \sim \Gamma^{-0.20}$. In R-IV ($\Gamma \gtrsim 20$), the effect of vertical confinement vanishes, resulting in a vertical stack of rolls. The flow approaches an asymptotic channel-like state in which $Nu$, $Re_u$, and $Re_v$ ($Re_u/Re_v \approx 0.07$) become independent of $\Gamma$. 

Resolvent analysis reveals that, with $Nu'$ as the quantity of interest, the maximum gain drops from R-I to R-II owing to a stable LSC that is the least sensitive to perturbations. However, the gain increases abruptly in regimes R-III and R-IV, indicating a strong sensitivity of the LSC, which could be responsible for the breakdown of a single LSC into multiple rolls. 
We also note that the most amplified response mode shifts from a diagonally dominant structure peaking at the adiabatic walls in R-I to a uniform, bulk-dominated structure in R-II. In regimes R-III and R-IV, the response mode takes the form of a vertical stack of rolls.

A key result is the identification of an optimal transport pathway controlled by LSC anisotropy. Across all $Ra$, the heat flux is maximized when $Re_u/Re_v \approx 0.45$, which defines a balance between horizontal heat-carrying motion and buoyancy-driven vertical flow. The corresponding optimal geometry follows the scaling law $\Gamma_{\mathrm{opt}} \sim Ra^{-0.17\pm0.01}$. This criterion provides a physically interpretable framework for geometry-driven heat-transfer optimization, with direct implications for cavity design, and flow-control strategies. 

\section*{Acknowledgment}
The authors thank Prof.\ K.~R.~ Sreenivasan for insightful suggestions on the  significance of Reynolds number in DHC, and Dr. M. Sharma for insightful suggestions. 
\appendix
\section{Simulation details}\label{app:1}

\begin{table}\label{tab:1}
\centering
\scriptsize

\begin{minipage}[t]{0.5\textwidth}
\centering
\setlength{\tabcolsep}{8pt}
\caption{Simulation parameters for $Ra=10^6$ with $N_x=128$. From left: Case number, aspect ratio $(\Gamma)$,  number of grid points in the $y$ direction  ($N_y$), Nusselt number ($Nu$), flow state (S and U denote steady and unsteady states, respectively), and length of sampling in time ($\tau_{\text{avg}}$).}
\label{tab:1}
\begin{tabular}{cccccc}
\toprule
Case No. & $\Gamma$ & $N_y$ & $Nu$ & State & $\tau_{\text{avg}}$ \\
\midrule
1  & 0.10   & 32   & 1.02  & S & --   \\
2  & 0.215  & 32   & 3.35  & S & --   \\ 
3  & 0.25   & 32   & 4.06  & S & --   \\ 
4  & 0.28   & 48   & 5.55  & S & --   \\
5  & 0.3    & 48   & 6.52  & S & --   \\
6  & 0.4    & 64   & 7.02  & S & --   \\
7  & 0.50   & 64   & 8.33  & S & --   \\
8  & 0.60   & 80   & 8.79  & S & --   \\
9  & 0.70   & 90   & 8.96  & S & --   \\
10 & 0.80   & 104  & 9.01  & S & --   \\
11 & 1      & 128  & 8.98  & S & --   \\
12 & 1.215  & 160  & 8.84  & S & --   \\
13 & 1.464  & 200  & 8.64  & S & --   \\
14 & 2      & 256  & 8.39  & S & --   \\
15 & 3      & 384  & 7.91  & S & --   \\
16 & 4      & 512  & 7.27  & S & --   \\
17 & 5      & 640  & 6.86  & U & 2000   \\
18 & 6      & 768  & 6.56  & U & 2000 \\
19 & 8      & 1024 & 6.32  & U & 2000 \\
20 & 10     & 1280 & 6.01  & U & 2000 \\
21 & 13     & 1600 & 5.74  & U & 2000 \\
22 & 16     & 2048 & 5.22  & U & 2000 \\
23 & 20     & 2560 & 5.24  & U & 1000 \\
24 & 30     & 3840 & 5.32  & U & 1000 \\
25 & 40     & 5120 & 5.32  & U & 1000 \\
26 & 50     & 6400 & 5.33  & U & 400 \\
27 & 60     & 7680 & 5.33  & U & 400 \\
\bottomrule
\end{tabular}
\end{minipage}
\hfill
\begin{minipage}[t]{0.48\textwidth}
\centering
\setlength{\tabcolsep}{8pt}
\caption{Simulation parameters for $Ra=10^5,3\times 10^6, 10^7,$ and $10^8$ with $N_x=64,160,256,$ and $512$ respectively.}
\label{tab:2}
\begin{tabular}{ccccc}
\toprule
S. No. & $\Gamma$ & $N_y$ & $Nu$ & $Ra$ \\
\midrule
1 & 0.2      & 32    & 1.05310 & $10^5$ \\
2 & 0.5      & 32    & 3.78212 & $10^5$ \\
3 & 0.8      & 64    & 4.47265 & $10^5$ \\
4 & 1        & 64    & 4.53831 & $10^5$ \\
5 & 2        & 128   & 4.30940 & $10^5$ \\
6 & 10       & 640   & 3.16771 & $10^5$ \\
7 & 20       & 1280  & 2.69411 & $10^5$ \\
8 & 30       & 1920  & 2.4518  & $10^5$ \\
9 & 40       & 2560  & 2.47876 & $10^5$ \\
\hline
10 & 0.5      & 128  & 12.41  & $3\times10^6$ \\
11 & 0.6      & 160  & 12.46  & $3\times10^6$ \\
12 & 0.7      & 180  & 12.39  & $3\times10^6$ \\
13 & 0.8      & 208  & 12.26  & $3\times10^6$ \\
14 & 1        & 256  & 11.94  & $3\times10^6$ \\ \hline
15 & 0.2      & 64   & 14.61  & $10^7$ \\
16 & 0.4      & 128  & 17.79  & $10^7$ \\
17 & 0.5      & 128  & 17.84  & $10^7$ \\
18 & 0.6      & 160  & 17.66  & $10^7$ \\
19 & 0.7      & 180  & 17.41  & $10^7$ \\
20 & 0.8      & 208  & 17.12  & $10^7$ \\
21 & 1        & 256  & 16.54  & $10^7$ \\
22 & 2        & 512  & 14.40  & $10^7$ \\ \hline
23 & 0.2      & 128  & 34.431  & $10^8$ \\
24 & 0.3      & 192  & 35.361  & $10^8$ \\
25 & 0.4      & 256  & 34.924  & $10^8$ \\
26 & 0.5      & 256  & 34.110  & $10^8$ \\
27 & 0.6      & 320  & 33.234  & $10^8$ \\
28 & 0.7      & 360  & 32.393  & $10^8$ \\
29 & 0.8      & 416  & 31.612  & $10^8$ \\
30 & 1        & 512  & 30.238  & $10^8$ \\
\bottomrule
\end{tabular}
\end{minipage}
\end{table}

The governing equations \ref{eq:1}-\ref{eq:3} are discretized using FVM on a collocated grid framework, where second-order central difference scheme is used for spatial derivatives. Time integration is performed using a fractional-step method where the nonlinear convective terms are advanced explicitly using a second-order Adams-Bashforth scheme, while the diffusive and buoyancy terms are advances via a second-order Crank-Nicolson scheme for enhanced stability (CFL remains below $0.3$), see \citet{sharma_influence_2022,sharma_dominant_2024} for more details). 

\bibliographystyle{jfm}
\bibliography{bibgraphy,oldbibgraphy,references}

@article{Bejan1979,
  author    = {Bejan, A.},
  title     = {Note on {G}ill's solution for free convection in a vertical enclosure},
  journal   = {Journal of Fluid Mechanics},
  year      = {1979},
  volume    = {90},
  number    = {3},
  pages     = {561--568},
  doi       = {10.1017/S0022112079002391}
}

@article{Cormack1974a,
  author    = {Cormack, D. E. and Leal, L. G. and Imberger, J.},
  title     = {Natural convection in a shallow cavity with differentially heated end walls.
               {P}art~1. {A}symptotic theory},
  journal   = {Journal of Fluid Mechanics},
  year      = {1974},
  volume    = {65},
  number    = {2},
  pages     = {209--229},
  doi       = {10.1017/S0022112074001352}
}

@article{Cormack1974b,
  author    = {Cormack, D. E. and Leal, L. G. and Seinfeld, J. H.},
  title     = {Natural convection in a shallow cavity with
               differentially heated end walls.
               {P}art~2. {N}umerical solutions},
  journal   = {Journal of Fluid Mechanics},
  year      = {1974},
  volume    = {65},
  number    = {2},
  pages     = {231--246},
  doi       = {10.1017/S0022112074001364}
}

@article{Ostrach1988,
  author    = {Ostrach, S.},
  title     = {Natural convection in enclosures},
  journal   = {Journal of Heat Transfer},
  year      = {1988},
  volume    = {110},
  number    = {4b},
  pages     = {1175--1190},
  doi       = {10.1115/1.3250619}
}

@article{LeQuere1998,
  author    = {{Le Qu\'{e}r\'{e}}, P. and Behnia, M.},
  title     = {From onset of unsteadiness to chaos in a
               differentially heated square cavity},
  journal   = {Journal of Fluid Mechanics},
  year      = {1998},
  volume    = {359},
  pages     = {81--107},
  doi       = {10.1017/S0022112097008458}
}

@article{Howland2022,
  author    = {Howland, C. J. and Ng, C. S. and Verzicco, R. and
               Lohse, D.},
  title     = {Boundary layers in turbulent vertical convection at
               high {P}randtl number},
  journal   = {Journal of Fluid Mechanics},
  year      = {2022},
  volume    = {930},
  pages     = {A32},
  doi       = {10.1017/jfm.2021.940}
}

@article{Ravi_1994, title={On the high-Rayleigh-number structure of steady laminar natural-convection flow in a square enclosure}, volume={262}, DOI={10.1017/S0022112094000522}, journal={Journal of Fluid Mechanics}, author={Ravi, M. R. and Henkes, R. A. W. M. and Hoogendoorn, C. J.}, year={1994}, pages={325–351}}

@book{bejan2013convection,
  title={Convection Heat Transfer},
  author={Bejan, Adrian},
  edition={4th},
  year={2013},
  publisher={John Wiley \& Sons},
  address={Hoboken, NJ},
  isbn={978-0470900376}
}

@article{macgregor_free_1969,
    title = {Free {Convection} {Through} {Vertical} {Plane} {Layers}—{Moderate} and {High} {Prandtl} {Number} {Fluids}},
    volume = {91},
    issn = {0022-1481},
    url = {https://doi.org/10.1115/1.3580194},
    doi = {10.1115/1.3580194},
    number = {3},
    urldate = {2026-03-17},
    journal = {Journal of Heat Transfer},
    author = {MacGregor, R. K. and Emery, A. F.},
    month = aug,
    year = {1969},
    pages = {391--401},
}

@article{GL_2000,
	Author = {S. {Grossmann} and D. {Lohse}},
	Journal = {J. Fluid Mech.},
	Month = {November},
	Pages = {27-56},
	Title = {Scaling in thermal convection: a unifying theory},
	Volume = {407},
	Year = {2000},
}

@article{Ahlers_2009rev,
	Author = {G. {Ahlers} and S. {Grossmann} and D. {Lohse}},
	Journal = {Rev. Mod. Phys.},
	Month = {June},
	Pages = {503-537},
	Title = {Heat transfer and large scale dynamics in turbulent \uppercase{R}ayleigh-\uppercase{B}{\'e}nard convection},
	Volume = {81},
	Year = {2009},
}

@article{Ng_2015,
title={Vertical natural convection: application of the unifying theory of thermal convection},
volume={764},
DOI={10.1017/jfm.2014.712},
journal={Journal of Fluid Mechanics},
publisher={Cambridge University Press},
author={Ng, C. S. and Ooi, A. and Lohse, D. and Chung, D.},
year={2015},
pages={349-361}
}

@article{Wang_2019,
author = {Wang, Q.  and Wan, Z.-H.  and Yan, R.  and Sun, D.-J. },
title = {Flow organization and heat transfer in two-dimensional tilted convection with aspect ratio 0.5},
journal = {Physics of Fluids},
volume = {31},
number = {2},
pages = {025102},
year = {2019},
doi = {10.1063/1.5070132},
}

@article{Wang_2021, title={Regime transitions in thermally driven high-Rayleigh number vertical convection}, volume={917}, DOI={10.1017/jfm.2021.262}, journal={Journal of Fluid Mechanics}, publisher={Cambridge University Press}, author={Wang, Q. and Liu, H.-R. and Verzicco, R. and Shishkina, O. and Lohse, D.}, year={2021}, pages={A6}}

@article{Paolucci_Chenoweth_1989, title={Transition to chaos in a differentially heated vertical cavity}, volume={201}, DOI={10.1017/S0022112089000984}, journal={Journal of Fluid Mechanics}, author={Paolucci, Samuel and Chenoweth, Donald R.}, year={1989}, pages={379–410}}

@article{Gill1969,
  author = {Gill, A. E. and Davey, A.},
  title = {Instabilities of a buoyancy-driven system},
  journal = {Journal of Fluid Mechanics},
  year = {1969},
  volume = {35},
  pages = {775--798}
}

@article{jovanovic_componentwise_2005,
	title = {Componentwise energy amplification in channel flows},
	volume = {534},
	copyright = {https://www.cambridge.org/core/terms},
	issn = {0022-1120, 1469-7645},
	url = {https://www.cambridge.org/core/product/identifier/S0022112005004295/type/journal_article},
	doi = {10.1017/S0022112005004295},
	language = {en},
	urldate = {2026-08-17},
	journal = {Journal of Fluid Mechanics},
	author = {Jovanović, Mihailo R. and Bamieh, Bassam},
	month = jul,
	year = {2005},
	pages = {145--183},
}

@article{mckeon_critical-layer_2010,
	title = {A critical-layer framework for turbulent pipe flow},
	volume = {658},
	copyright = {https://www.cambridge.org/core/terms},
	issn = {0022-1120, 1469-7645},
	url = {https://www.cambridge.org/core/product/identifier/S002211201000176X/type/journal_article},
	doi = {10.1017/S002211201000176X},
	language = {en},
	urldate = {2026-08-17},
	journal = {Journal of Fluid Mechanics},
	author = {McKEON, B. J. and Sharma, A. S.},
	month = sep,
	year = {2010},
	pages = {336--382},
}

@article{rolandi_invitation_2024,
	title = {An invitation to resolvent analysis},
	volume = {38},
	issn = {1432-2250},
	url = {https://doi.org/10.1007/s00162-024-00717-x},
	doi = {10.1007/s00162-024-00717-x},
	language = {en},
	number = {5},
	urldate = {2026-04-06},
	journal = {Theoretical and Computational Fluid Dynamics},
	author = {Rolandi, Laura Victoria and Ribeiro, Jean Hélder Marques and Yeh, Chi-An and Taira, Kunihiko},
	month = oct,
	year = {2024},
	pages = {603--639},
}

@article{gill_boundary-layer_1966,
	title = {The boundary-layer regime for convection in a rectangular cavity},
	volume = {26},
	copyright = {https://www.cambridge.org/core/terms},
	issn = {0022-1120, 1469-7645},
	url = {https://www.cambridge.org/core/product/identifier/S0022112066001368/type/journal_article},
	doi = {10.1017/S0022112066001368},
	language = {en},
	number = {3},
	urldate = {2026-03-16},
	journal = {Journal of Fluid Mechanics},
	author = {Gill, A. E.},
	month = nov,
	year = {1966},
	pages = {515--536},
}

@article{sharma_dominant_2024,
	title = {Dominant heat transfer mechanism with conical roughness in a cubical box in turbulent {Rayleigh}–{Bénard} convection},
	volume = {36},
	issn = {1070-6631},
	doi = {10.1063/5.0206619},
	number = {6},
	urldate = {2024-06-29},
	journal = {Physics of Fluids},
	author = {Sharma, Mukesh and Chand, Krishan and De, Arnab Kr.},
	month = jun,
	year = {2024},
	pages = {065124},
}

@article{sharma_influence_2022,
	title = {Influence of {Prandtl} number in turbulent {Rayleigh}-{Bénard} convection over rough surfaces},
	volume = {7},
	issn = {2469-990X},
	doi = {10.1103/PhysRevFluids.7.104609},
	language = {en},
	number = {10},
	urldate = {2024-05-30},
	journal = {Physical Review Fluids},
	author = {Sharma, Mukesh and Chand, Krishan and De, Arnab Kr.},
	month = oct,
	year = {2022},
	pages = {104609},
}

@article{padovan_resolvent4py_2025,
	title = {Resolvent4py: {A} parallel {Python} package for analysis, model reduction and control of large-scale linear systems},
	volume = {31},
	issn = {23527110},
	shorttitle = {Resolvent4py},
	url = {https://linkinghub.elsevier.com/retrieve/pii/S2352711025002523},
	doi = {10.1016/j.softx.2025.102286},
	language = {en},
	urldate = {2025-08-18},
	journal = {SoftwareX},
	author = {Padovan, Alberto and Anantharaman, Vishal and Rowley, Clarence W. and Vollmer, Blaine and Colonius, Tim and Bodony, Daniel J.},
	month = sep,
	year = {2025},
	pages = {102286},
}

@article{chand_effect_2022,
	title = {Effect of inclination angle on heat transport properties in two-dimensional {Rayleigh}–{Bénard} convection with smooth and rough boundaries},
	volume = {950},
	issn = {0022-1120, 1469-7645},
	url = {10.1017/jfm.2022.815},
	doi = {10.1017/jfm.2022.815},
	language = {en},
	urldate = {2024-05-30},
	journal = {Journal of Fluid Mechanics},
	author = {Chand, Krishan and Sharma, Mukesh and De, Arnab Kr},
	month = nov,
	year = {2022},
	pages = {A16},
}

@article{shishkina_momentum_2016,
	title = {Momentum and heat transport scalings in laminar vertical convection},
	volume = {93},
	issn = {2470-0045, 2470-0053},
	url = {https://link.aps.org/doi/10.1103/PhysRevE.93.051102},
	doi = {10.1103/PhysRevE.93.051102},
	language = {en},
	number = {5},
	urldate = {2023-09-02},
	journal = {Physical Review E},
	author = {Shishkina, Olga},
	month = may,
	year = {2016},
	pages = {051102},
}
\end{document}